\documentclass[aps,pra,twocolumn,showpacs,superscriptaddress]{revtex4-1}
\usepackage{graphicx,amsmath,amssymb}
\usepackage[usenames]{color}
\begin{document}

\title{Ground-state phase diagram of the quantum Rabi model}

\author{Zu-Jian Ying}
\email{Electronic address: zjying@csrc.ac.cn}
\affiliation{Beijing Computational Science Research Center, Beijing 100084, China}
\affiliation{CNR-SPIN, I-84084 Fisciano (Salerno), Italy and Dipartimento di Fisica ``E. R. Caianiello", Universit$\grave{a}$  di Salerno, I-84084 Fisciano (Salerno), Italy}

\author{Maoxin Liu}
\affiliation{Beijing Computational Science Research Center, Beijing 100084, China}

\author{Hong-Gang Luo}
\email{Electronic address: luohg@lzu.edu.cn}
\affiliation{Center for Interdisciplinary Studies $\&$ Key Laboratory for Magnetism and Magnetic Materials of the MoE,
Lanzhou University, Lanzhou 730000, China}
\affiliation{Beijing Computational Science Research Center, Beijing 100084, China}

\author{Hai-Qing Lin}
\email{Electronic address: haiqing0@csrc.ac.cn}
\affiliation{Beijing Computational Science Research Center, Beijing 100084, China}

\author{J. Q. You }
\affiliation{Beijing Computational Science Research Center, Beijing 100084, China}

\begin{abstract}
The Rabi model plays a fundamental role in understanding light-matter interaction. It reduces to the Jaynes-Cummings model via the rotating-wave approximation, which is applicable only to the cases of near resonance and weak coupling. However, recent experimental breakthroughs in upgrading light-matter coupling order require understanding the physics of the full quantum Rabi model (QRM). Despite the fact that its integrability and energy spectra have been exactly obtained, the challenge to formulate an exact wavefunction in a general case still hinders physical exploration of the QRM. Here we unveil a ground-state phase diagram of the QRM, consisting of a quadpolaron and a bipolaron as well as their changeover in the weak-, strong- and intermediate-coupling regimes, respectively. An unexpected overweighted antipolaron is revealed in the quadpolaron state, and a hidden scaling behavior relevant to symmetry breaking is found in the bipolaron state. An experimentally accessible parameter is proposed to test these states, which might provide novel insights into the nature of the light-matter interaction for all regimes of the coupling strengths.
\end{abstract}
\maketitle

\textit{Introduction.--} In the past decade, it has been witnessed that the exploration of fundamental quantum physics in light-matter coupling systems has significantly evolved toward the (ultra-)strong coupling regime \cite{Wallraff2004, Gunter2009, Niemczyk2010, Peropadre2010, Forn-Diaz2010, Cristofolini2012, Scalari2012, Xiang2013}. For example, in 2004, the strong coupling of a single microwave photon to a superconducting qubit was realized experimentally by using circuit quantum electrodynamics \cite{Wallraff2004}. In 2010, this coupling rate was enhanced to reach a considerable fraction up to 12\% of the cavity transition frequency\cite{Niemczyk2010}. Even with such small fractions the system has already entered into so-called ultrastrong-coupling limit\cite{Ciuti2006,Devoret2007}. In this situation, the well-known Jaynes-Cummings (JC) model\cite{JC1963} is no longer applicable because the JC model is valid only in the cases of near resonance and weak coupling \cite{Allen1987}. Indeed, the experimentally observed anticrossing in the cavity transmission spectra \cite{Niemczyk2010} was due to counter-rotating terms, which are dropped in the JC model as a rotating-wave approximation. In addition, experimental observation of the Bloch-Siegert shift \cite{Forn-Diaz2010} also requires taking into account the counter-rotating terms in the description of the JC model. Thus the importance of the counter-rotating terms raises requests to comprehend the behavior of a full quantum Rabi model \cite{rabi1936, rabi1937, Cohen1992} (QRM) for all regimes of the coupling strengths \cite{Ciuti2005, Bourassa2009, Casanova2010, Irish2014}.

Remarkably, an important progress in the study of the QRM in the past years is the proof of its integrability \cite{Braak2011, Solano2011}. As a result, its energy spectra have been exactly obtained \cite{Braak2011, Chen2012}. However, to calculate the dynamics of the system, correlation
functions, and even other simpler physical observables, it is not enough to know only the exact eigenvalues, but the wavefunctions
(e.g., the exact eigenstates) are desirable. Based on series expansions of the eigenstates in terms of known basis sets, it was realized that a standard calculation with double precision, sufficient to compute the spectrum, fails for the eigenstates\cite{Wolf2012}. Therefore, the challenge to formulate an exact wave function in a general case still hampers  access to a full understanding of the QRM.

In this work, by deforming the polaron and antipolaron \cite{Bera2014a,Bera2014b} we propose a novel variational wavefunction ansatz to extract the ground state physics of the QRM. It is found that this ansatz is valid with high accuracy in all regimes of the coupling strengths. Thus a ground state phase diagram of the QRM is constructed. The nature of the system variation, by increasing the coupling strength from weak to strong, becomes transparent in the ground-state phase diagram with a quantum state changeover from quadpolaron to bipolaron, around a novel critical-like coupling scale analytically extracted. In particular, an unexpected overweighted antipolaron is revealed in the quadpolaron state and a hidden scaling behavior is found in the bipolaron state. Moreover, we propose an experimentally accessible parameter to test these states. For perspective, we also extend this ansatz to the multiple-mode case, which is expected to be useful to understand the physics of the spin-boson model \cite{Spin-Boson}.

\begin{figure}[h]
\begin{center}
\includegraphics[width=0.9\columnwidth]{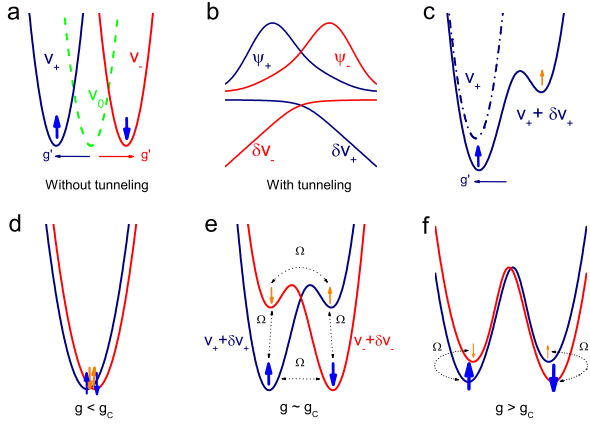}
\end{center}
\caption{(Color) Schematic diagram for effective potentials induced by the tunneling between two levels.
\textbf{a}, In the absence of tunneling, i.e., $\Omega = 0$, the
original harmonic oscillator ($v_0$) is coupled with two
levels denoted by $\uparrow$ and $\downarrow$ to form two
polarons (associated with $v_\pm$) with the left($+,\uparrow$) and
right($-,\downarrow$) displacement $g'=\sqrt{2}g/ \omega$.
\textbf{b}, When the tunneling $\Omega$ is switched on, the left-
and right-polarons provide an effective
potential for each other $\delta v_\pm = \eta  {\frac \Omega {  \omega}}  \frac
{\psi_\mp}{\psi_\pm}$ ($\eta = \pm$ represents the parity, here we
focus on the ground state with ``$-$" parity), which induces an
antipolaron, as shown in c. \textbf{c}, The potential of the
left-polaron deforms from $v_+$ to $v_+ + \delta v_+$. The size of
$\uparrow$ indicates the weight of the polaron (blue) and
antipolaron (orange), respectively, in the same and opposite
directions of the potential displacement. The situation is
symmetric for the right-polaron. \textbf{d}-\textbf{f}, Typical
deformed potentials in the weak ($g < g_c$), intermediate ($g \sim g_c$), and strong
coupling ($g > g_c$) cases. There exist four tunneling channels between $\uparrow$ and
$\downarrow$ states, as shown in \textbf{e}, forming a \textit{quadpolaron} state. In the strong coupling case, the tunneling
between left and right states decays until it is vanishingly small due to the large
potential barriers between them, yielding a \textit{bipolaron}
state in \textbf{f}.} \label{fig1}
\end{figure}

\textit{The model and effective potential.--} The QRM \cite{rabi1936, rabi1937} describes a quantum two-level
system coupled to a single bosonic mode or quantized harmonic
oscillator, which is a paradigm for interacting quantum systems
ranging from quantum optics \cite{cavityQED} to quantum information
\cite{raimond} to condensed matter \cite{holstein}. The model
Hamiltonian reads
\begin{equation}
H= \omega a^{\dagger }a +\frac \Omega 2 \sigma_x +
g\sigma_z(a^{\dagger }+a), \label{ham}
\end{equation}
where $a^\dagger(a)$ is the bosonic creation (annihilation) operator
with frequency $\omega$ and $\sigma_{x,z}$ is the Pauli matrix
with level splitting $\Omega$.  The last term describes the
interaction with coupling strength $g$.

In terms of the quantum harmonic oscillator with dimensionless
formalism\cite{Mahan2000} $a^{\dagger}=\left(\hat x-i\hat
p\right)/\sqrt{2}$, $a=\left( \hat x+i\hat p\right)/\sqrt{2}$,
where $\hat x=x$ and $\hat p=-i\frac \partial {\partial x}$ are
the position and momentum operators, respectively, 
the model can be rewritten as
\begin{equation}
H =  \sum_{\sigma_z=\pm } \left(h^{\sigma _z}|\sigma _z\rangle
\langle \sigma _z|+ {\frac \Omega 2} |\sigma _z\rangle \langle
\overline{\sigma}_z|\right) + {\cal E}_0,  \label{Hx}
\end{equation}
where $\overline{\sigma }_z = -\sigma _z$ and $+$($-$) labels the
up $\uparrow$ (down $\downarrow$) spin in the $z$ direction,
respectively. $h^{\pm}=\frac12 \omega (\hat p^2+ v_\pm)$, with
$v_\pm = \left( \hat x \pm g'\right)^2$ and $g'=\sqrt{2}g/\omega$,
while ${\cal E}_0=-\frac12  \omega (g'^2 + 1)$ is a constant energy. Apparently, $h^{\pm}$ define two bare polarons \cite{Bera2014a, Bera2014b} in the sense that the harmonic oscillator is bound by $\sigma_z$ due to the coupling $g'$, as shown in Fig. \ref{fig1}a. These two polarons form two bare potential wells but the existence of the level splitting $\Omega$ (resulting in the tunneling between these two wells\cite{Irish2014}) makes the model difficult to solve analytically.

Let us begin with the wave-function $\Psi$ satisfying the Schr\"odinger equation $H\Psi = E\Psi$ with the eigenenergy $E$. Due to the fact that the model possesses the parity symmetry, namely, $[{\cal P},H]=0$ with ${\cal P}=\sigma_x(-1)^{a^{\dagger }a}$, $\Psi$ should take the form of
$\Psi = \frac{1}{\sqrt{2}}\left(\psi_{+}|\uparrow\rangle + \eta \psi_{-}|\downarrow\rangle\right)$,
where $\psi _{\pm} = \psi(\pm x)$ will be given below and $\eta = 1$ $(-1)$ for
positive (negative) parity. Without loss of generality, here we consider the ground state, with negative parity. The Schr\"odinger equation becomes
\begin{equation}
\frac12   \omega (\hat p^2 + v_\pm + \delta v_\pm )\psi_\pm = E\psi_\pm,
\end{equation}
where $\delta v_\pm = - \frac \Omega { \omega }\frac
{\psi_\mp}{\psi_\pm}$ is an additional effective potential originating from the tunneling, as shown in the lower
panel of Fig. \ref{fig1}b. The additional potential will deform the bare potential and as a result creates a subwell in the opposite direction of the bare potential $v_{\pm}$, as illustrated in Fig. \ref{fig1}c. The subwell induces an \textit{antipolaron} as a
quantum effect. The above analysis from potential subwell verifies the existence of
antipolaron from wavefunction expansion\cite{Bera2014a, Bera2014b}. Thus, the polaron and antipolaron constitute the basic ingredients of the ground-state wavefunction.

\textit{Deformed polaron and antipolaron.--}
With the concept of polaron and antipolaron in hand, the competition between different energy scales $\omega, \Omega$ and $g'$ involved in the QRM will inevitably lead to deformations of the polaron and antipolaron. Physically, they can deform predominantly in two possible ways:
the position is shifted and the frequency is renormalized, which will introduce four independent variational parameters given below.
Explicit deformation depends on the coupling strength once the
tunneling is fixed, as shown in Fig. \ref{fig1} d-f from weak to
strong couplings according to a critical-like coupling strength
$g_c$. Thus a trial variational wave-function for $\psi(x)$
takes the superposition of the deformed polaron ($\varphi_\alpha$) and
antipolaron ($\varphi_\beta$),
\begin{equation}
\psi(x)=\alpha \varphi_\alpha(x) + \beta \varphi_\beta(x),
\label{wf-1}
\end{equation}
where $\varphi_\alpha (x)= \phi_0\left( \xi_\alpha \omega, x +
\zeta_\alpha g'\right)$ and $\varphi_\beta (x)= \phi_0\left(
\xi_\beta \omega,  x - \zeta_\beta g'\right)$, with
$\phi_0(\omega, x)$ being the ground-state of standard harmonic oscillator with
frequency $\omega$. Here $\xi_i$ ($\zeta_i$), with $i = \alpha$
and $\beta$, describes the renormalization for frequency
(displacement) independently for the polaron and the antipoalron,
while the coefficients of $\alpha$ and $\beta$ denote their
weights, subject to the normalization condition
$\langle\psi|\psi\rangle = 1$. We stress that in contrast to the
direct expansion on basis series without frequency renormalization
\cite{Bera2014a, Bera2014b}, we design our trial wavefunction based
on the dominant physics of deformation.

\begin{figure}[t]
\begin{center}
\includegraphics[width=1.0\columnwidth]{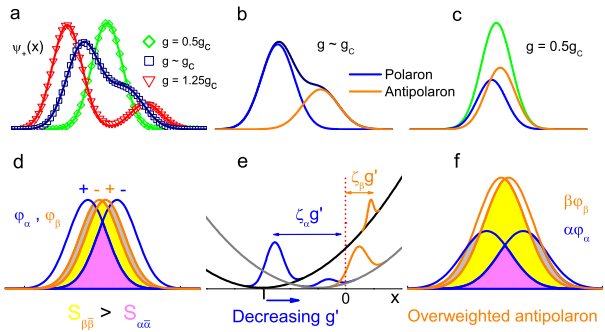}
\end{center}
\vskip0cm \caption{(Color) Mechanism for overweighted antipolaron in quadpolaron state. \textbf{a}, The calculated
(solid lines) spin-up groundstate wavefunctions, $\psi_+(x) =
\alpha\varphi_\alpha(x) + \beta \varphi_\beta(x)$, at $g/g_c = 0.5,
1., 1.25$ respectively, for weak (green), intermediate(navy),
and strong(red) couplings, with $\omega/\Omega = 0.1$. The symbols
denote the numerical exact results. The spin-down wavefunction is
given by $-\psi_+(-x)$(not shown). \textbf{b} and \textbf{c},
$\alpha$- and $\beta$-components of $\psi_+(x)$, which correspond
to the polaron(blue) and antipolaron(orange), respectively, for
the intermediate $g \sim g_c$ and weak $g = 0.5g_c$ coupling
cases. \textbf{d}, The overlaps between different polarons and/or
antipolarons without the weights, $S_{i\bar{j}}=\langle
\varphi_i(x)|\varphi_j(-x)\rangle$ with $i,j = \alpha,\beta$. It
is clear that $S_{\beta\bar{\beta}}(\text{yellow}) >
S_{\alpha\bar{\alpha}}(\text{light magenta})$. \textbf{e},
Schematic illustration of the physics for the overweighted
antipolaron. When decreasing the coupling strength $g'$, 
the potential provided by the left-displaced oscillator for the
antipolaron gets lower, so the tunneling energy gain from large
$S_{\beta\bar{\beta}}$ in \textbf{d} overwhelms the potential
cost, which favors a larger weight of antipolaron. \textbf{f}, The
overweighted antipolaron with a larger weight than the polaron. }
\label{fig-Mechanism}
\end{figure}

It turns out that our variational wavefunction is capable of
providing a reliable analysis on the QRM in the whole parameter regime,
ranging from weak to strong couplings, as shown for several
physical quantities for the ground-state including the energy, the mean
photon number, the coupling correlation and the tunneling
strength in Appendix \ref{App-A}. Obviously the
remarkable agreement between our results and the exact ones roots
in the fact that our trial wavefunction correctly captures the basic physics, as illustrated by the accurate wavefunction profiles
compared to the exact numerical ones for various couplings in
Fig.\ref{fig-Mechanism}a. The variational wavefunction, with its
concise physical ingredients and its accuracy, in turn facilitates
unveiling more underlying physics.

\textit{Quadpolaron/bipolaron quantum state changeover.--}  From
Fig.\ref{fig-Mechanism}a-c one sees that when increasing the coupling, the wavepacket splits into visible polaron and
antipolaron (see animated plots in Supplementary Material for more vivid evolutions of
potentials and wavepackets). Before the full splitting, there are
significant tunnelings in all the four channels between the
polarons and antipolarons, as schematically shown in Fig.
\ref{fig1}e. Thus, in this sense we call this state a \textit{quadpolaron}. After the splitting, only two same-side channels of
tunneling survives while the left-right channels are blocked gradually by
the increasing barrier, as sketched in Fig. \ref{fig1}f. This
state is termed here as a \textit{bipolaron}. Despite the evolution from a transition-like feature in the low frequency limit to a crossover behavior in finite frequencies for the changover between quadpolaron and bipolaron states, the nature of the afore-mentioned splitting is essentially the same. This enables us to obtain an analytic coupling scale (see Appendix \ref{App-B}), $g_c = \sqrt{\omega^2 + \sqrt{\omega^4 + g^4_{c0}}}$, which generalizes the low frequency-limit result\cite{Ashhab2010} $g_{c0}=\sqrt{\omega\Omega}/2$ and correctly captures the quantum state changeover between quadpolaron and bipolaron for the whole range of frequencies.

\begin{figure}[t]
\begin{center}
\includegraphics[width=0.9\columnwidth]{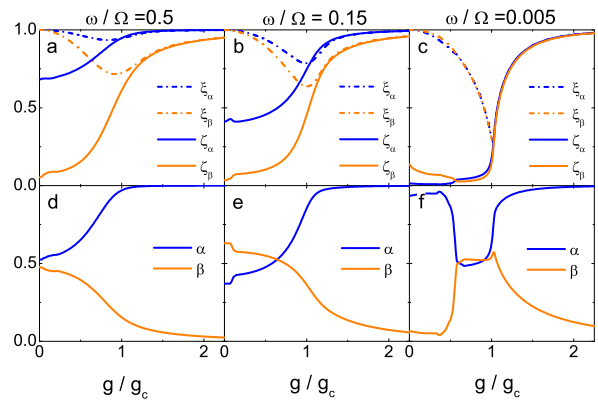}
\end{center}
\caption{(Color) Renormalization factor and weight as a function of the coupling strength $g$. $\zeta _i$ ($i=\alpha,\beta$) is the displacement renormalization and $\xi _i$ is the frequency renormalization. $\alpha$ and $\beta$ denote the weights of the polaron and the antipolaron, respectively, in the variational groundstate wavefunction.
\textbf{a} and \textbf{d}, $\omega/\Omega = 0.5$.
\textbf{b} and \textbf{e}, $\omega/\Omega = 0.15$.
\textbf{c} and \textbf{f}, $\omega/\Omega = 0.005$. }
\label{fig-Parameters}
\end{figure}
\textit{Quadpolaron asymmetry and overweighted antipolaron in the regime of $g \lesssim g_c$.--} We find that the polaron and antipolaron in the quadpolaron state have asymmetric displacements, which leads to a subtle competition depending on the frequency $\omega/\Omega$. Figure \ref{fig-Parameters} shows three types of distinct behaviors of the variational parameters in three different frequency regimes: high frequency ($\omega/\Omega \gtrsim 0.47$), intermediate frequency($\omega/\Omega \in [0.07, 0.47]$), and low frequency ($\omega/\Omega \lesssim 0.07$). The result is understandable due to the fact that the antipolaron always has a higher potential energy owing to its opposite direction to the displacement of $v_\pm$. Roughly speaking, at a high frequency, the antipolaron should have a lower weight than the polaron ($\beta < \alpha$) since the antipolaron is suppressed by the high potential. At a low frequency, the polaron benefits from both potential and tunneling energies. However, competition becomes subtle at an
intermediate frequency as each of these different energy scales may only favor either the polaron or the antipolaron respectively,
which may lead to overweighted antipolaron, as shown in Fig.\ref{fig-Parameters}e.

Below we give a more explicit analysis. Actually, the four channel
tunneling energies in the quadpolaron are proportional to the
overlaps of the polarons and antipolarons, $S_{\alpha
\bar{\alpha}}$, $S_{\beta \bar{\beta}}$, $S_{\alpha \bar{\beta}}$
and $S_{\beta \bar{\alpha}}$, respectively, as shown in
Fig.\ref{fig-Mechanism}d. The mixture terms $S_{\alpha
\bar{\beta}}$ and $S_{\beta \bar{\alpha}}$ do not affect the
weight competition between the polaron and antipolaron, while
$S_{\alpha \bar{\alpha}}$ and $S_{\beta \bar{\beta}}$ yield
imbalances. Indeed, the antipolarons have larger overlap than the
polarons, i.e. $S_{\beta\bar{\beta}}
> S_{\alpha\bar{\alpha}}$ (see Fig. \ref{fig-Mechanism}d). This is because the
antipolarons in up and down spins are closer to each other than
the polarons in order to reduce their higher potential energy, as indicated
in Fig. \ref{fig-Mechanism}e and quantitatively shown by $ \zeta _\beta < \zeta
_\alpha$ in Fig. \ref{fig-Parameters}a and \ref{fig-Parameters}b.
Therefore, as far as the tunneling is concerned,
it would tend to have more weight of antipolarons to gain a maximum tunneling.
When the intermediate frequency reduces the cost of potential energy for such tendency,
a larger antipolaron weight might finally occur, as in Fig. \ref{fig-Mechanism}f,
leading to an unexpected overweighted antipolaron. We find that this really
occurs as demonstrated in Fig. \ref{fig-Parameters}e where a
weight reversion appears at the crossing of $\alpha$ and $\beta$
for a weaker coupling.

At the low frequency, the harmonic potential becomes very flat,
the polarons may get closer than antipolarons, as indicated by
$\zeta _\alpha < \zeta _\beta$ in Fig.\ref{fig-Parameters}c in the
weak coupling regime.  In this case, $S_{\alpha \bar{\alpha} }$ is
greater than $S_{\beta \bar{\beta} }$ so that polarons have
favorable energies in both potential and tunneling. Thus the
polaron regains its priority in weight.

\begin{figure}[t]
\begin{center}
\includegraphics[width=0.9\columnwidth]{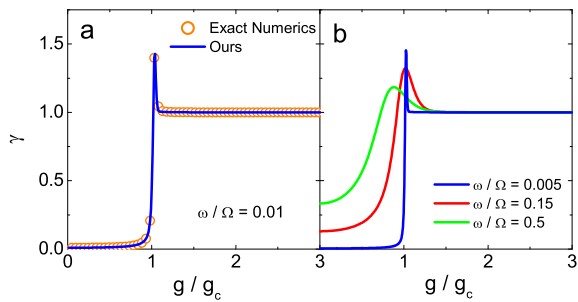}
\end{center}
\vskip-0.cm \caption{(Color) Scaling quantity $\gamma$ as a function of the coupling strength $g$. \textbf{a}, Our results
compared with exact numerics at $\omega/\Omega=0.01$ as an
example. \textbf{b}, $\gamma$ for different values of the ratios
$\omega/\Omega$. The scaling relation $\zeta_i \doteq \xi_i$ is
tested by $\gamma=1$ beyond $g_c$. } \label{fig-Scaling}
\end{figure}
\textit{Bipolaron and hidden scaling behavior in the regime of $g\gtrsim g_c$.--}
In the bipolaron regime, the remaining tunneling in channels $S_{\alpha\bar\beta}$ and $S_{\beta\bar\alpha}$ leads to intriguing physics,
showing a deeper nature of the interaction in the symmetry breaking aspect.
Indeed, Fig.\ref{fig-Parameters}a-c show that in this regime the frequency factor $\xi _i$ and the displacement factor $\zeta _i$ collapse into the same value,
i.e $\zeta_i \doteq \xi_i$. In fact, due to vanishing photon number below $g_c$ at low frequency limit, the parity ${\cal P}$ can be decomposed into
separate spin and spatial reversal sub-symmetries which are broken beyond $g_c$. However, further seeking the symmetry breaking character from these
sub-symmetries would fail at finite frequencies due to emergence of a finite number of photons below $g_c$. Nevertheless, the $\zeta_i$-$\xi_i$
symmetric aspect revealed here provides a compensation, from the beyond-$g_c$ side instead but valid also for finite frequencies.
To test this scaling behavior, we propose an experimentally accessible quantity, $\gamma
\equiv \frac \omega {gt} \sqrt{
 \langle a^{+}a \rangle -\frac 1 4 (t+t^{-1})+\frac 12
      }$,
where $t=-\langle \left( a^{+}-a\right) ^2\rangle $, which becomes the scaling ratio $\gamma \rightarrow \zeta /\xi$ (see Appendix \ref{App-C}) for their
average $\xi =\left( \xi _\alpha+\xi _\beta\right) /2$  and $\zeta =\left( \zeta _\alpha+\zeta _\beta\right) /2$
and thus equals to one above $g_c$, as shown in Fig. \ref{fig-Scaling} for various frequencies. The experimental measurement of $\gamma$ thus
provides a possible way to distinguish the states of bipolaron and quadpolaron as well as their changeover.

\begin{figure}[h]
\begin{center}
\includegraphics[width=0.95\columnwidth]{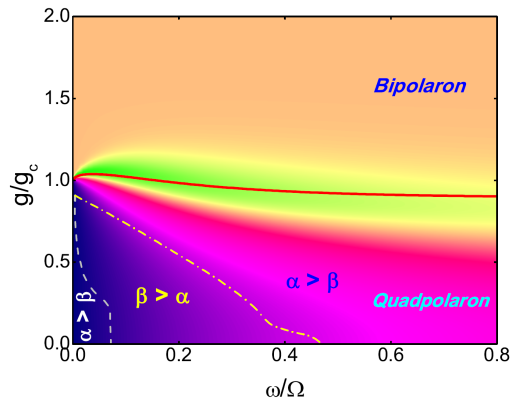}
\end{center}
\vskip-0.0cm \caption{(Color) An overview of ground-state phase diagram for the QRM. Quadpolaron ($g \lesssim g_c$) and bipolaron ($g \gtrsim g_c$) as well as their crossover near the minimum of $\xi_i$ (red solid line) or the maximum of $\gamma$ around analytic $g_c$. The quadpolaron regime is further divided into the normal ($\alpha > \beta$) and overweighted antipolaron ($\alpha < \beta$) regimes. The dashed and dot-dashed lines have been obtained numerically from the cross points as shown in Fig. \ref{fig-Parameters} (e) and (f). The color density for $\gamma$ further distinguishes the characters of the different regimes and their changeovers.} \label{fig-Diagram}
\end{figure}

\textit{Ground-state phase diagram.--} The above discussions on
polaron-antipolaron competition can be summarized into a ground-state phase
diagram unveiled, as shown in Fig.
\ref{fig-Diagram}. The ground-state with different channels of tunneling is identified as a quadpolaron when $g \lesssim g_c$ and as a bipolaron when $g\gtrsim g_c$. An overweighted antipolaron is hidden in the quadpolaron
regime, while a scaling relation between the displacement and frequency renormalizations is revealed in the
bipolaron regime.
Note that the polaron and antipolaron structures might be detected by
optomechanics\cite{Restrepo2014, Meystre2013} and $\gamma$ is experimentally
measurable. The diagram may provide a renewed panorama for deeper
theoretical investigations and may raise more challenges for experiments.

\textit{Perspective in multiple modes.--}
The basic physics in the QRM has a profound implication for the spin-boson model\cite{Spin-Boson}, which is a multiple-mode version of the QRM.
The essential variational ingredients remain similar. The trial wavefunction can be written as
 $
\psi [\{x_k\}]=\alpha \prod_{k=1}^M\varphi _\alpha^k+\beta \prod_{k=1}^M\varphi _\beta^k
 $
with the extension $\{\omega,g,x,\xi_i,\zeta_i\}\rightarrow
\{\omega_k,g_k,x_k,\xi^k_i,\zeta^k_i\}$ for the $k$'th mode. We illustrate the same accuracy by a two-mode case in Appendix \ref{App-D}.

\textit{Acknowledgements.--}
We thank Jun-Hong An for helpful discussions. Work at CSRC and Lanzhou University was supported by National Natural Science Foundation of China, PCSIRT (Grant No. IRT1251), National ``973" projects of China. Z.-J.Y. also acknowledges partial financial support from the Future and Emerging Technologies (FET)
programme within the Seventh Framework Programme for Research of the European Commission, under FET-Open grant number: 618083 (CNTQC).

\appendix

\section{Variational method and physical properties}\label{App-A}

Here we calculate the ground-state physical properties
from the variational method, including the energy $E$, the mean photon number $\langle a^{\dagger }a\rangle $,
the coupling correlation $\langle \sigma _z(a^{\dagger }+a)\rangle $ and
the spin flipping (tunneling) strength $\langle \sigma _x\rangle $.

\subsection{The energy}

As introduced in the main part of the paper, the wavefunction for the reformulated
Hamiltonian \eqref{Hx} has the
following form
\begin{equation}
\Psi =\frac 1{\sqrt{2}}\left( \psi _{+}(x)\mid \uparrow \rangle
+\eta \psi _{-}(x)\mid \downarrow \rangle \right) ,
\end{equation}
where $\eta =\pm $ is the parity. We adopt the variational trial wavefunction as a superposition of
the polaron and the antipolaron
\begin{equation}
\psi _{+}(x)=\psi _{-}(-x)=\alpha \varphi _\alpha \left( x\right)
+\beta \varphi _\beta \left( x\right) , \label{wf-1}
\end{equation}
where
\begin{eqnarray}
\varphi _\alpha \left( x\right) =\phi _n(\xi _\alpha \omega,x+\zeta _\alpha g^{\prime }),\\
\varphi _\beta \left( x\right) =\phi _n(\xi _\beta \omega ,x-\zeta _\beta g^{\prime }),
\end{eqnarray}
with $\phi _n(\omega, x)$ being the $n$'th eigenstate of the
standard quantum harmonic oscillator with frequency $\omega $. In
this work we focus on the ground state so that $n=0$ and $\eta
=-$. The displacement of the bare potential $v_\pm = \left( \hat x \pm g'\right)^2$ in the single-well energy $ h^{\pm },$
\begin{equation}
g^{\prime }=\sqrt{2}g/\omega ,
\end{equation}
is driven by the interaction $g$ and for simplicity we have assumed
the unit $\hbar = m = 1$. Note that the polaron (antipolaron) has a displacement in the
same (opposite) direction as (to) that of the bare potential
$v^{\pm }$. The interplay of the interaction and the tunneling
leads to the deformation of the wavepacket: the frequency of the
polaron (antipolaron) is renormalized by $\xi _\alpha $ ($\xi
_\beta $) and the displacement by $\zeta _\alpha $ ($\zeta _\beta
$), respectively. The weights of the polaron and the antipolaron are
subject to the normalization condition $\langle \Psi \left| \Psi
\right\rangle =\langle \psi _{+}|\psi _{+}\rangle =1$. These
deformation parameters, independently $\{\xi _\alpha ,\xi _\beta
,\zeta _\alpha ,\zeta _\beta ,\alpha \}$, are optimized by
minimization of the total energy formulated in the following.

The energy can be directly obtained as
\begin{equation}
E\equiv \left\langle \Psi \right| H\left| \Psi \right\rangle
=h_{++}^{+}+\eta \frac{\hbar \Omega }2n_{+-}+{\cal E}_0,
\label{energy-1mode}
\end{equation}
where
\begin{eqnarray}
h_{++}^{+}&=&\left\langle \psi _{+}\right| h^{+}\left| \psi
_{+}\right\rangle \nonumber \\
&=&\alpha ^2h_{\alpha \alpha }^{+}+\beta
^2h_{\beta \beta }^{+}+2\alpha \beta h_{\alpha \beta }^{+}, \label{hpp}\\
n_{+-}&=&\langle \psi _{+}|\psi _{-}\rangle \nonumber \\
&=&\alpha ^2S_{\alpha
\bar \alpha }+\beta ^2S_{\beta \bar \beta }+2\alpha \beta
S_{\alpha \bar \beta }, \label{nab}
\end{eqnarray}
contribute to the single-well energy and the tunneling energy,
respectively. Here, we have defined
\begin{eqnarray}
h_{ij}^{+}&=&\langle \varphi _i\left( x\right) |h^{+}|\varphi
_j\left( x\right) \rangle , \nonumber \\
S_{ij}&=&\langle \varphi _i\left(
x\right) \mid \varphi _j\left( x\right) \rangle ,\label{defSij} \\
S_{i\,\bar j}&=&\langle \varphi _i\left( x\right) \mid \varphi _j\left(
-x\right) \rangle ,  \nonumber
\end{eqnarray}
for $i=\alpha ,\beta $, while ${\cal E}_0=-\frac 12 \omega
(1+g^{\prime 2})$ is a constant energy. Explicit formulas for
these quantities are readily available. In this Section we give the result
for the ground state
\begin{eqnarray}
h_{\alpha \alpha }^{+} &=&\frac 12\omega \left[ \frac 12(\xi
_\alpha +\xi
_\alpha ^{-1})+\left( 1-\zeta _\alpha \right) ^2g^{\prime 2}\right] , \label{hAaa}\\
h_{\beta \beta }^{+} &=&\frac 12\omega \left[ \frac 12(\xi _\beta
+\xi
_\beta ^{-1})+\left( 1-\zeta _\beta \right) ^2g^{\prime 2}\right] , \label{hAbb}\\
h_{\alpha \beta }^{+} &=&\frac 12\omega
\left[ \left( 1-\xi _\alpha ^2\right) \left\langle \hat
x_\alpha ^2\right\rangle _{\alpha \beta } +\left( 1-\zeta _\alpha
\right) \left\langle \hat x_\alpha \right\rangle _{\alpha \beta
}2g^{\prime } \right. \nonumber\\
&& \left.
+\xi _\alpha S_{\alpha \beta }+\left( 1-\zeta _\alpha \right) ^2g^{\prime
2}S_{\alpha \beta }\right] ,\label{hAab}
\end{eqnarray}
where
\begin{eqnarray}
\left\langle \hat x_\alpha \right\rangle _{\alpha \beta }
&=&S_{\alpha\beta} \frac{\left( \zeta _\alpha +\zeta _\beta
\right) \xi _\beta }{\left( \xi
_\alpha +\xi _\beta \right) }g^{\prime },  \label{x1Aab} \\
\left\langle \hat x_\alpha ^2\right\rangle _{\alpha \beta }
&=&\frac {S_{\alpha\beta}} {(\xi _\alpha +\xi _\beta )}\left[  1+\frac{\left( \zeta _\alpha +\zeta
_\beta \right)  ^2\xi _\beta ^2}{(\xi _\alpha +\xi _\beta
)}g^{\prime 2}\right]
\label{x2Aab}
\end{eqnarray}
and
\begin{eqnarray}
S_{\alpha \beta }&=&S(\zeta _\alpha
,\zeta _\beta ,\xi _\alpha
,\xi _\beta ),
\nonumber \\
S_{\alpha \bar \beta }&=&S(\zeta _\alpha ,-\zeta
_\beta ,\xi _\alpha ,\xi _\beta ),
\nonumber  \\
S_{\alpha \bar \alpha } &=& S(\zeta _\alpha ,\zeta _\alpha ,\xi
_\alpha ,\xi _\alpha ),
\nonumber \\
S_{\beta \bar \beta } &=& S(\zeta _\beta ,\zeta _\beta ,\xi _\beta
,\xi _\beta ),
\end{eqnarray}
are given by the function
\begin{eqnarray}
S(\zeta _1,\zeta _2,\xi _1,\xi _2) &=&  \exp \left(
-\frac{\left( \zeta _1+\zeta _2\right) ^2g^{\prime 2}\xi _1\xi
_2}{2(\xi _1+\xi _2)}\right)    \nonumber \\
&& \times \sqrt{2} \left[ \frac{\xi _1\xi _2}{(\xi
_1+\xi _2)^2}\right] ^{1/4}.
\end{eqnarray}

\subsection{The mean photon number}

From the relation
\begin{equation}
a^{\dagger }a=\frac{h^0}\omega -\frac 12,\quad h^0\equiv \frac
12\omega \left( \hat p^2+\hat x^2\right) ,
\end{equation}
and the symmetric relation $\psi _{-}\left( x\right) =\psi
_{+}\left( -x\right) $, the mean photon number simply reads as
\begin{equation}
\langle a^{\dagger }a\rangle \equiv \left\langle \Psi \right|
a^{\dagger }a\left| \Psi \right\rangle =\frac{h_{++}^0}\omega
-\frac 12  \label{PhotonN}
\end{equation}
where
\begin{equation}
h_{++}^0=\left\langle \psi _{+}\right| h^0\left| \psi
_{+}\right\rangle =\alpha ^2h_{\alpha \alpha }^0+\beta ^2h_{\beta
\beta }^0+2\alpha \beta h_{\alpha \beta }^0.
\end{equation}
For the ground state
\begin{eqnarray}
h_{\alpha \alpha }^0 &=&\frac 12\omega \left[ \frac 12(\xi _\alpha
^{-1}+\xi
_\alpha )+2\zeta _\alpha ^2g^{\prime 2}\right] , \\
h_{\beta \beta }^0 &=&\frac 12\omega \left[ \frac 12(\xi _\beta
^{-1}+\xi
_\beta )+2\zeta _\beta ^2g^{\prime 2}\right] , \\
h_{\alpha \beta }^0 &=&\frac 12\omega \left[ \left( 1-\xi _\alpha ^2\right) \left\langle \hat x_\alpha
^2\right\rangle _{\alpha \beta }-\zeta _\alpha \left\langle \hat
x_\alpha \right\rangle _{\alpha \beta }2g^{\prime } \right.\nonumber \\
&&\left.+\xi _\alpha S_{\alpha
\beta }+\zeta _\alpha
^2g^{\prime 2}S_{\alpha \beta }\right] ,
\end{eqnarray}
and $\left\langle \hat x_\alpha ^2\right\rangle _{\alpha \beta }$,
$ \left\langle \hat x_\alpha \right\rangle _{\alpha \beta }$ are
given by \eqref{x1Aab} and \eqref{x2Aab}.

\subsection{The coupling correlation $\langle \sigma _z(a^{\dagger
}+a)\rangle $ and the spin flipping (tunneling) strength $\langle
\sigma _x\rangle $}

Now we calculate the coupling correlation $\langle \sigma
_z(a^{\dagger }+a)\rangle $. Since $(a^{\dagger }+a)=\sqrt{2}\hat
x$, we have
\begin{equation}
\langle \sigma _z(a^{\dagger }+a)\rangle \equiv \left\langle \Psi
\right| \sigma _z(a^{\dagger }+a)\left| \Psi \right\rangle
=\sqrt{2}\left\langle \hat x\right\rangle _{++}
\end{equation}
where
\begin{equation}
\left\langle \hat x\right\rangle _{++}=\left\langle \psi
_{+}\left( x\right) \right| \hat x\left| \psi _{+}\left( x\right)
\right\rangle =\alpha ^2x_{\alpha \alpha }+\beta ^2x_{\beta \beta
}+2\alpha \beta x_{\alpha \beta }
\end{equation}
and
\begin{equation}
x_{\alpha \alpha }=-\zeta _\alpha g^{\prime },\quad x_{\beta \beta
}=\zeta _\beta g^{\prime },\quad x_{\alpha \beta }=\left\langle
x_\alpha \right\rangle _{\alpha \beta }-\zeta _\alpha S_{\alpha
\beta }g^{\prime }.
\end{equation}

The strength of spin flipping or tunneling, $\sigma _x=\sigma
^{+}+\sigma ^{-}$, is simply
\begin{eqnarray}
\langle \sigma _x\rangle \equiv \left\langle \Psi \right| \sigma
_x\left| \Psi \right\rangle =\eta n_{+-}
\end{eqnarray}
which has been formulated in \eqref{nab}.

\subsection{Accuracy of our variational method}

\begin{figure}[t]
\begin{center}
\includegraphics[width=0.9\columnwidth]{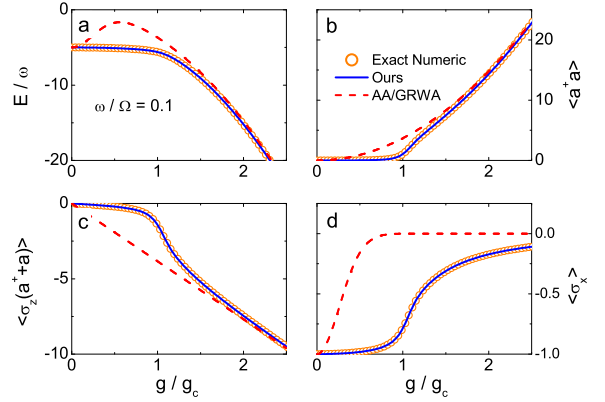}
\end{center}
\par
\caption{(Color online) Ground-state physical quantities as functions of the coupling strength $g/g_c$. $\omega/\Omega = 0.1$ is taken as an example. \textbf{a}, The ground state
energy. \textbf{b}, The mean photon number. \textbf{c}, The correlation
function $\langle \sigma_z (a^\dagger + a)\rangle$. \textbf{d}, The
tunneling strength $\langle \sigma_x\rangle$. The orange circles
denote the numerically exact results as a benchmark, the red
dashed-lines are calculated in adiabatic approximation(AA)
\cite{Irish2005} or generalized rotating-wave approximation (GRWA)
\cite{Irish2007}, and the blue lines are our results obtained by
the present variational method.} \label{fig-PhyQuan}
\end{figure}

The most widely-used approximations in the literature are the
rotating-wave approximation (RWA)\cite{JC1963}, adiabatic
approximation(AA) \cite{Irish2005}, generalized rotating-wave
approximation (GRWA) \cite{Irish2007} and generalized variational
method (GVM) \cite{GVM1, GVM2}, each working in some specific parameter
regime. The RWA neglects the counter-rotating terms in the
interaction, valid in regime $g\ll \omega ,\Omega $ under
near-resonance ($ \omega \sim \Omega $) condition. The AA and the
GRWA have the same groundstate, working for $g\gg \omega $ or
negative detuning ($\omega >\Omega $) regime. The GVM works for
$g\ll \omega $. Recently a mean-photon-number dependent
variational method was proposed to cover validity regimes
of both the GVM and the GRWA \cite{LiuM}. However, all the
approximations collapse when the ratio of $\omega /\Omega $ is
getting small, e.g. below around $0.5$ (see Ref. [\onlinecite{LiuM}]). An improved
variational method by including the antipolaron\cite{Irish2014}
also finds breakdowns at $\omega /\Omega \sim 0.3$. It would be
favorable to have a variational method that always preserves a
high accuracy in varying all parameters which might facilitate and
even deepen the physical analysis.

Indeed, our variational wavefunction yields such accuracy
requirements. As an illustration, in Fig. \ref{fig-PhyQuan} we
compare with the exact numerics on the the ground state energy,
mean photon number, coupling correlation and tunneling strength,
at the example $\omega /\Omega =0.1$ (one can find other examples
for comparison at $\omega /\Omega =0.01$, $0.05$, $0.15$, $0.5$ for
another physical quantity $\gamma $ in Fig. \ref{fig-Scaling} and Fig. \ref{fig-Sij}). As a comparison, the
results obtained by the AA or the GRWA are also shown. Clearly,
our results are completely consistent with the exact ones in the
whole parameter regime.

\subsection{Physical necessity of the variational parameters}

\begin{figure}[tbp]
\begin{center}
\includegraphics[width=0.9\columnwidth]{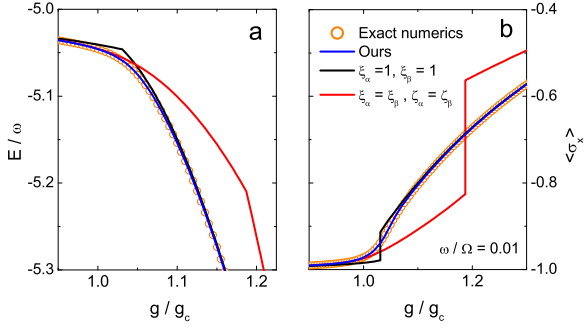}
\end{center}
\par
\caption{(Color online) Quantitative deviations and qualitative errors emerge in reducing variational parameters. Physical properties may deviate not only quantitatively but also qualitatively when the parameters are reduced, e.g., if imposing $\xi _\alpha=1,\ \xi _\beta=1$ (black line) or $\xi _\alpha=\xi _\beta, \ \zeta _\alpha=\zeta _\beta$ (red line),  an incorrect cusp behavior appears in the energy $E$ and the spin flipping (tunneling) strength $\langle \sigma _x\rangle $ has a spurious jump around $g_c$ at $\omega/\Omega = 0.01$, in contradiction with the smooth crossover in the exact numerics (orange circles). The blue lines are our results in full minimal parameters which reproduce accurately the exact ones.  }
\label{fig-ParamtNecessity}
\end{figure}
\begin{figure}[tbp]
\begin{center}
\includegraphics[width=0.8\columnwidth]{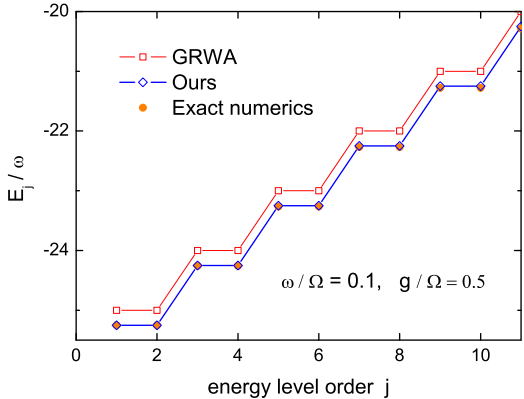}
\end{center}
\par
\caption{(Color online) An energy comparison of the excited states for the lowest
10 levels, at $\omega/\Omega = 0.1$ and $g/\Omega = 0.5$. The
orange dots, empty blue diamonds and empty red squares represent
the results of the exact numerics, our method and the GRWA,
respectively. } \label{fig-Excited}
\end{figure}
It may be worthwhile to have further discussion on the physical necessity of the variational parameters. An unnecessary reduction of our parameters, on the one hand, will not lead to much of a reduction in the computational cost as the calculation in full parameters is actually quite easy to carry out, on the other hand, however, the price of physical loss would be too high. As discussed in the main part of the paper, our variational parameters physically correspond to the deformations of the polaron and the antipolaron with displacement and frequency renormalizations, which is justified by the behavior of the effective potential.  In the subtle energy competitions of the potential of harmonic oscillator, the interaction and the tunneling, both the polaron and the antipolaron can adapt themselves via the variations of their displacements, frequencies and weights. Thus, corresponding to the key physical degree of freedom of the polaron and the antipolaron, the five variational parameters, $\xi _\alpha,\ \xi _\beta, \ \zeta _\alpha,\ \zeta _\beta, \alpha $,  are the minimal necessary parameters to capture the true physics of the behavior of the polaron and the antipolaron, subject to the normalization of the wavefunction. Therefore, reducing the parameters would lead to mismatch of the physical degree of freedom and thus give rise to unreliable results, the physical properties may deviate not only quantitatively but also qualitatively. For an example, assuming $\xi _\alpha=\xi _\beta=1$ or imposing $\xi _\alpha=\xi _\beta, \ \zeta _\alpha=\zeta _\beta$ can reduce the parameter number by 2. However, as shown in Fig.\ref{fig-ParamtNecessity}, without mentioning the quantitative deviations, an incorrect cusp behavior appears in the energy $E$ at low frequencies as illustrated at $\omega/\Omega = 0.01$, and even worse, a spurious jump emerges in the tunneling (spin flipping) strength $\langle \sigma _x\rangle $ around $g_c$.  The other physical quantities, such as the mean photon number $\langle a^{\dagger }a\rangle $, the coupling correlation $\langle \sigma _z(a^{\dagger }+a)\rangle $ also have a false discontinuity similar to $\langle \sigma _x\rangle $.  Both the cusp and the discontinuity are qualitatively in contradiction with the smooth crossover in the exact numerics (orange circles). Nevertheless, our results using the full minimal variational parameters (blue line) reproduce accurately the exact results in the entire regime of the coupling strengths at different frequencies. Moreover, in the cases of reduced parameters, some important underlying physics would also be missing, such as the scaling relation of the displacement and frequency renormalizations as we revealed in the main text (see also Appendix \ref{App-C}).

\subsection{Method extension to the excited states}

Our method can also be useful for the excited states. As a first simple extension the variational energy of the excited
state can be obtained by replacing expressions \eqref{hAaa}-\eqref{hAab} with
\begin{eqnarray}
h_{\alpha \alpha }^{+} &=&\frac \omega 2\left[ \left( n+\frac
12\right) \left( \xi _\alpha +\xi _\alpha ^{-1}\right) +\left(
1-\zeta _\alpha \right)
^2g^{\prime 2}\right] , \\
h_{\beta \beta }^{+} &=&\frac \omega 2\left[ \left( n+\frac
12\right) \left( \xi _\beta +\xi _\beta ^{-1}\right) +\left(
1-\zeta _\beta \right)
^2g^{\prime 2}\right] , \\
h_{\alpha \beta }^{+} &=&\frac \omega 2\ \left[
\left( 1-\xi _\alpha ^2\right) \left\langle \hat
x_\alpha ^2{}\right\rangle _{\alpha \beta }+
\left( 1-\zeta _\alpha
\right) \left\langle \hat x_\alpha \right\rangle _{\alpha \beta
}2g^{\prime }\right. \nonumber \\
&&+ \left. (2n+1)\xi _\alpha
S_{\alpha \beta }+
\left( 1-\zeta _\alpha \right) ^2g^{\prime
2}S_{\alpha \beta }\right],
\end{eqnarray}
where both $\langle \hat x_\alpha ^j{}\rangle _{\alpha \beta }$ and $%
S_{\alpha \beta }$ can be included by a unified function
\begin{equation}
\langle \hat x_\alpha ^j{}\rangle _{\alpha \beta }=X(n,j),\quad
S_{\alpha \beta }=X(n,0).
\end{equation}
\begin{widetext}
Here the function $X(n,j)$ is defined by
\begin{eqnarray}
X(n,j) &=&n!j!\left[\frac{\left( \zeta _\alpha +\zeta _\beta \right) g^{\prime }}{%
2c}\right]^j\sum_{p=0}^{\min [n,j]}\sum_{q=0}^{\min [n,j-p]}\frac{\left(
-i\right) ^{j-p-q}a^pb^q}{p!q!\left( j-p-q\right)
!}\sqrt{\frac{2^{p+q}}{\left( n-p\right) !\left( n-q\right)
!}}H_{j-p-q}\left(\frac 12ab^2c\right)\tilde S_{n-p,n-q},
\\
\tilde S_{k,k^{\prime }} &=&\sum_{r=0}^{\min [k,k^{\prime
}]}C_{kk^{\prime
}r}H_{k-r}\left(\frac{ab^2c}{2\sqrt{1-a^2}}\right)H_{k^{\prime }-r}\left(-\frac{a^2bc}{2%
\sqrt{1-b^2}}\right), \\
C_{kk^{\prime }r} &=&\sqrt{\frac{ab}{2^{k+k^{\prime }}k!k^{\prime }!}}%
e^{-\left( abc\right) ^2/4}\frac{k!k^{\prime }!\left( 2ab\right)
^r\left( 1-a^2\right) ^{(k-r)/2}\left( 1-b^2\right) ^{(k^{\prime
}-r)/2}}{\left( k-r\right) !\left( k^{\prime }-r\right) !r!}.
\end{eqnarray}
and the factors $a,b,c$ depend on the variational parameters
\begin{equation}
a=\sqrt{\frac{2\xi _\alpha }{\xi _\alpha +\xi _\beta }},\quad b=\sqrt{\frac{%
2\xi _\beta }{\xi _\alpha +\xi _\beta }},\quad c=\left( \zeta
_\alpha +\zeta _\beta \right) g^{\prime }\sqrt{\frac{\left( \xi
_\alpha +\xi _{\beta}\right) }2}.
\end{equation}
\end{widetext}

For the other group of overlap in the tunneling term $n_{+-}$\eqref{nab}, 
one can also formulate using $S_{\mu \bar \mu ^{\prime }}=\left(
-1\right) ^nX(n,0)$ with
the corresponding replacement $\alpha ,\beta \rightarrow \mu ,\mu ^{\prime }$, but there is sign variation $\zeta _\beta \rightarrow -\zeta
_{\mu ^{\prime }}$. Here $n$ is the level number of the standard
quantum harmonic oscillator and $H_m\left( x\right) $ is the
standard Hermite polynomials. It is worthwhile to see that this
simple extension for the excited states has already yielded some
considerable improvements in strong couplings as illustrated in
Fig.\ref{fig-Excited} for a number of lowest energy levels. With
the above expressions, one may further analytically construct an
improved extension of the variational energy for overall coupling
range by imposing the deformed polaron and antipolaron in the GRWA
form of wavefunction. On the other hand, the dynamics of the system also can be calculated in terms of $\tilde S_{k,k^{\prime}}$ which provides the intra-overlap and inter-overlap of the deformed polarons and antipolarons with different oscillator quantum number $k,k^{\prime }$.
Since here the focus is the ground state
which, as we show in the present work, already has rich underlying
physics to be uncovered, we shall present a more detailed method
description and systematical discussion for the excited state
properties in our future work.

\section{Quadpolaron/bipolaron changeover and scales of coupling strength}\label{App-B}

\subsection{Analytic approximation for $g_c$}

In the variation of the coupling strength, the system undergoes a
phase-transition-like changeover around $g\sim g_c$. In the
super-strong tunneling or low-frequency limit, i.e. $\omega
/\Omega \rightarrow 0,$ this changeover is very sharp, it behaves
more like a phase transition, as discussed by Ashhab
\cite{Ashhab2013}. In the other cases it behaves like a crossover.

\begin{figure}[t]
\begin{center}
\includegraphics[width=1.0\columnwidth]{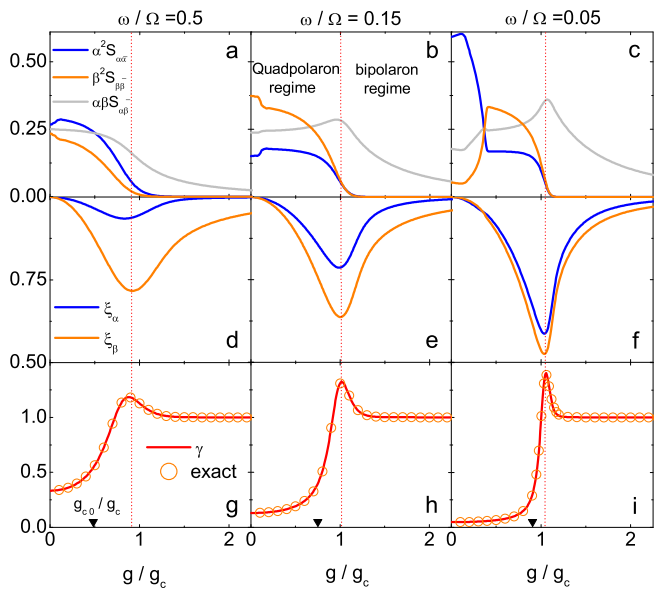}
\end{center}
\par
\vskip+0.0cm \caption{(Color online) Quadpolaron/bipolaron changeover and the behavior of variational parameters and related physical quantities. \textbf{a-c}, Weighted
groundstate tunneling of different channels, $\alpha ^2 S_{\alpha
\bar{\alpha}}$, $\beta ^2 S_{\beta \bar{\beta}}$, $\alpha \beta
S_{\alpha \bar{\beta}}$ and $\alpha \beta S_{\beta \bar{\alpha}}$
($S_{\beta \bar{\alpha}}= S_{\alpha \bar{\beta}}$) as functions of
the coupling strength $g/g_c$. The dashed lines roughly separate
the quadpolaron ($g\lesssim g_c$) regime and the bipolaron regime
($ g \gtrsim g_c$), the former has four channels of tunnelings
while the latter has two channels. \textbf{d-f}, The frequency
renormalization factors $ \xi_\alpha$ and $\xi_\beta$. \textbf{g-i},
The scaling quantity $\gamma$. The results from our variational
method (solid lines) almost reproduce the ones from exact numerics
(orange circles) for all values of coupling at different
frequencies. The boundary of the quadpolaron and bipolaron regimes
is associated with the minimum of $\xi_i$ and the maximum of
$\gamma$. The black triangles mark the positions for $g_{c0}/g_c$
which becomes farther away from $1$ as $\omega$ increases. \textbf{a,d} $\omega/\Omega = 0.5$. \textbf{b, e},
$\omega/\Omega =0.15$. \textbf{c,f}, $\omega/\Omega = 0.05$. }
\label{fig-Sij}
\end{figure}

We can get more insights into this transition-like behavior from
the profile deformation of the wavepacket. The increase of the
coupling strength is splitting the wavepacket into the polaron and the
antipolaron, while the tunneling is trying to keep them as close
as possible in the groundstate. Before a full splitting the system
remains in a quadpolaron state with four channels of tunneling,
$S_{\alpha \bar \alpha }$, $S_{\beta \bar \beta }$, $ S_{\alpha
\bar \beta }$, and $S_{\beta \bar \alpha }$, while after the
splitting the system enters a bipolaron state with only two
tunneling channels, $S_{\alpha \bar \beta }$ and $S_{\beta \bar
\alpha }$, surviving. Here we have labeled the tunneling channels
by the overlaps $S_{i\bar{j}}$, defined in \eqref{defSij}, to which the corresponding tunneling energies are proportional.
We show the tunneling channel difference for
these two regimes in Fig.\ref {fig-Sij} a-c at various
frequencies. One can also see that the change in the tunneling
channel number is universal for different frequencies. Thus, the
two regimes distinguished by wavepacket splitting are essentially
different in the quantum states. Therefore, the coupling strength
at which the splitting really starts can be used to formulate
$g_c$.

We adopt the value of $g_c$ at the point where the distance
between the polaron and the antipolaron is equal to their total radii,
\begin{equation}
\left( \zeta _\alpha +\zeta _\beta \right) g_c^{\prime }=r_\alpha
+r_\beta , \label{eqR12r12}
\end{equation}
where we take the radii by
\begin{equation}
r_\alpha =2\sqrt{\frac 1{\xi _\alpha }},\quad r_\beta
=2\sqrt{\frac 1{\xi _\alpha }},
\end{equation}
at which the value of the corresponding wavepacket is becoming
small
\begin{equation}
\frac{\varphi _i}{\varphi_i^{\max }}=\frac 1{e^2}
\end{equation}
for both $i=\alpha, \beta $.

Note that both sides of the above equation \eqref{eqR12r12} are
essentially averaging over the polaron and the antipolaron, thus
assuming symmetric polaron and antipolaron, i.e. $\zeta _\alpha
=\zeta _\beta $ and $\xi _\alpha =\xi _\beta $, would be a reasonable
approximation as far as $g_c$ is concerned. Under this constraint
the explicit results for the deformation parameters are available
for the well-separated polaron and antipolaron from the energy
minimization formulated in Appendix \ref{App-A}, reading as
\begin{equation}
\zeta _\alpha =\zeta _\beta =\sqrt{1-\frac{g_{c0}^4}{g^4}},\quad
\xi _\alpha =\xi _\beta =1,  \label{Xi-noSqueez}
\end{equation}
where the critical point $g_{c0}=\sqrt{\omega \Omega }/2$ is
obtained in the semiclassical approximation at $\omega /\Omega
\rightarrow 0.$\cite {Ref-gc0,Ashhab2010} We stress that we limit
the application of this approximation to the estimation of $g_c$,
while for other properties one should fall back upon asymmetric
polaron and antipolaron for higher accuracy. Actually, as mentioned in Appendix \ref{App-A}, imposing
symmetric polaron and antipolaron would lead to a spurious
discontinuous behavior of physical properties, such as the tunneling strength, around $g_c$ at
low frequencies, while in reality it
should be smooth as predicted by asymmetric polaron and
antipolaron in agreement with exact numerics.
Also, in the strong
coupling regime the displacement asymmetry of the polaron and the
antipolaron actually plays an important role in inducing the amplitude-squeezing effect ($\xi _\alpha <1$) which extends the wavepackets
of the polaron and the antipolaron to increase their overlap, thus enhancing the tunneling.
Without the asymmetry there would be no squeezing beyond $g_c$, as indicated by \eqref{Xi-noSqueez}, since the symmetric polaron and antipolaron in up and down spins would completely coincide, with an already-maximum overlap. In fact, as uncovered in the main text, there is a hidden relation between the squeezing and the displacement, which is also discussed in detail in Appendix \ref{App-C}.

Substitution of \eqref{Xi-noSqueez} into \eqref{eqR12r12} leads us
to a simple analytic expression
\begin{equation}
g_c=\sqrt{\omega ^2+\sqrt{\omega ^4+g_{c0}^4}}.  \label{eq-gC}
\end{equation}
It is easy to check $g_c\rightarrow g_{c0}$ in the slow oscillator
limit $ \omega /\Omega \rightarrow 0$. Besides the transition-like changeover in
this low frequency limit, our $g_c$ is also providing a valid
coupling scale for the quadpolaron/bipolaron changeover at finite
frequencies, which can be seen from Fig.\ref{fig-Sij} where the
quadpolaron regime and bipolaron regime adjoin each other really
around $g_c$. A more quantitative way to identify the
transition-like point is, as shown by Fig.\ref{fig-Sij} \textbf{d-i}, the
minimum point of the frequency renormalization factor or the
maximum point of the scaling quantity introduced in Appendix \ref{App-C}.
Still, one sees that it is well approximated by $g_c$ in
\eqref{eq-gC}.

\subsection{Novel scale for the coupling strength}

\begin{figure}[t]
\begin{center}
\includegraphics[width=1.0\columnwidth]{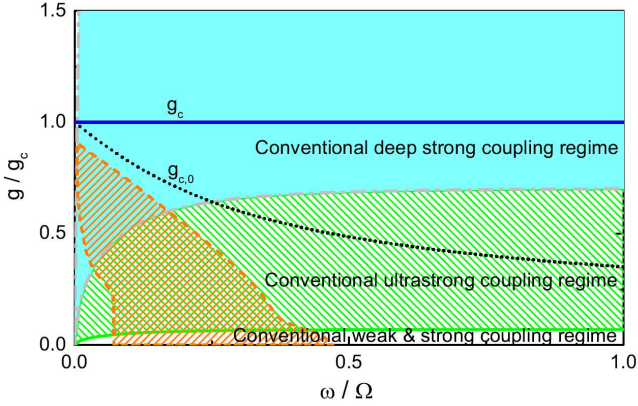}
\end{center}
\par
\vskip+0.0cm \caption{(Color online) The conventional coupling regimes used in the literature. The conventional ultrastrong coupling regime (the green
shaded area) is enclosed by $g=0.1\omega$ and $g=\omega$, which
has been reached by experiments in rapid
progress\cite{Niemczyk2010}. The conventional deep strong coupling
regime (the light-cyan area) is surrounded by $g=\omega$ and
$g=10\omega$ (gray dash-dot lines), into which investigations have been
entering\cite{Casanova2010}. The black dotted line denotes the
semiclassical critical-like point in low frequency limit,
$g_{c0}$\cite{Ref-gc0,Ashhab2010}, proposed as a different scale
of coupling strength\cite{Irish2014}, while the blue solid line
schematically represents the quadpolaron/bipolaron boundary $g_c$ as a novel scale
generalized for the whole range of frequencies. Thus, the coupling strength is divided into weak, intermediate and strong regimes which correspond to that $g$ is smaller than, comparable to and larger than $g_c$, respectively. The orange-shaded
window edged by the dash lines opens for the overweighted
antipolaron discussed in our paper.} \label{fig-different-scales}
\end{figure}

At this point, it is worthwhile to further discuss 
the scale of the coupling strength, the criterion for which is
actually a bit controversial in the literature\cite{Irish2014}.
Although the terms for the coupling strengths were given in
relation to the validity of the RWA as well as the progress of
experimental accessibility,  essentially the frequency $\omega $
has been conventionally taken as the evaluation scale: $ g/\omega \leq 0.01$ for
the weak coupling regime, $g/\omega \geq 0.01$ for the strong
coupling, $g/\omega \geq 0.1$ for the ultrastrong coupling
regime\cite{Niemczyk2010}, $g/\omega \geq 1$ for deep strong
coupling regime\cite{Casanova2010}. On the other hand, it should
be noticed that recently it has been proposed \cite{Irish2014}
that the strength scale should be modified to be the semiclassical
critical point $g_{c0}$. Still, as afore-mentioned, $g_{c0}$ is
obtained in low-frequency semiclassical limit, while the situations at finite
frequencies would be different. The controversy essentially comes from the fact that a consensus on the nature of the interaction-induced variation in different frequencies is still lacking. Here, our expression of $g_c$ in
\eqref{eq-gC} is obtained by the observation that it is the
wavepacket splitting that makes the essential change in increasing
the coupling strength, which controls the final effective coupling
tunneling strength and leads to transition (in low frequency
limit) or crossover (at finite frequencies) of the
quadpolaron/bipolaron states. We believe that $g_c$ is a more
universal scale valid for all frequencies, as indicated by Fig.
\ref{fig-Sij}. Under these
considerations, we simply divide the coupling
strength into weak, intermediate and strong regimes under the conditions that $g$ is smaller than, comparable to and larger
than $g_c$, respectively. As a reference, we compare the different scales for the coupling strength used in the literature in Fig.\ref{fig-different-scales}.

\section{Hidden scaling relation and  symmetry-breaking-like aspect }\label{App-C}

\subsection{Scaling relation extracted from energy minimization}\label{Sect-Scaling}

When the coupling strength grows beyond $g_c$, we find that
the squeezing factor $\xi _i$ and the displacement factor $\zeta
_i$ begin to collapse into the same values and scale with each other in
the further evolution, i.e.
\begin{equation}
\xi _\alpha \doteq \zeta _\alpha ,\quad \xi _\beta \doteq \zeta
_\beta . \label{scalingEq}
\end{equation}
This hidden scaling relation can be more explicitly formulated at
low frequencies. Note that the parameters can be extracted from
the energy minimization
\begin{equation}
\frac{\delta E}{\delta \alpha }=0,\quad \frac{\delta E}{\delta \xi
_i} =0,\quad \frac{\delta E}{\delta \zeta _i}=0.
\end{equation}
In the bipolaron regime, only the polaron-antipolaron tunneling
remains so the overlaps  $S_{\alpha \bar{\alpha}}$ and
$S_{\beta \bar{\beta}}$  are vanishing, but  $S_{\beta
\bar{\alpha}}$ is finite. In such a situation, controlling the
polaron-antipolaron center of mass, $ \zeta =\left( \zeta _\alpha
+\zeta _\beta \right) /2$, can be decoupled from the relative
motion in tunneling and squeezing, which enables us to extract
the weight of the polaron,
\begin{equation}
\alpha =\sqrt{\frac{1+\zeta _\beta }{2-\left( \zeta _\alpha -\zeta
_\beta \right) }}.  \label{alpha}
\end{equation}
To obtain analytical results we assume a low frequency which enables a small-%
$\omega $ expansion and leads us to
\begin{eqnarray}
\xi _{\alpha ,\beta }&=&\xi \left( 1\pm \frac{\omega
^2}{4g^2}\right) ,\nonumber \\
\zeta _{\alpha ,\beta }&=&\zeta \left( 1\pm
\frac{\omega ^2}{4g^2\left( 1-g_{c0}^4/g^4\right) }\right) ,
\end{eqnarray}
where $\xi _\alpha $ ($\xi _\beta $) takes the sign $+$ ($-$). In
the small-$ \omega $ limit, $\xi _i$ and $\zeta _i$ collapse into
their average $\xi =\left( \xi _\alpha +\xi _\beta \right) /2$ and
$\zeta =\left( \zeta _\alpha +\zeta _\beta \right) /2$ which are
equal:
\begin{equation}
\xi =\zeta =\sqrt{1-\frac{g_{c0}^4}{g^4}},
\end{equation}
up to $\omega ^2$ order. We can see the scaling relation from the
approximate analytic results: (i) in low-frequency limit, one sees
that $\xi _i\doteq \zeta _i$ holds, up to an $\omega ^2$
order correction which is negligible for small $\omega $. (ii) For
higher frequencies, the $\omega ^2$ terms in $\xi _{\alpha ,\beta
}$ and $\zeta _{\alpha ,\beta }$ become almost the same due to
$g_{c0}^4/g^4\ll 1$, since in bipolaron regime we have
$g>g_c>g_{c0}$ (e.g., for $\omega =0.5\Omega $,
$g_{c0}^4/g_c^4=0.056$ while $g_{c0}^4/g^4$ is negligible beyond
the crossover range.). These analytic considerations account for
the scaling relation as we showed in the main text for different
frequencies.

To test the scaling relation, we shall propose a physical
quantity that may be either measured experimentally or verified by
exact numerics. On one hand, applying the above expansion to the
photon number \eqref{PhotonN} and neglecting the difference of
$\xi _\alpha ,\zeta _\alpha $ and $\xi _\beta ,\zeta _\beta $
leads us to an expression of $\zeta $ as a function of $
\langle a^{+}a\rangle $ and $\xi $%
\begin{equation}
\zeta \doteq \frac \omega g\sqrt{\langle a^{+}a\rangle -\frac
14(\xi +\xi ^{-1})+\frac 12}.
\end{equation}
On the other hand, the same approximation yields
\begin{equation}
\xi \doteq -\langle \left( a^{+}-a\right) ^2\rangle .
\end{equation}
Thus, considering the ratio $\zeta /\xi $, we introduce the
following phyiscal quantity
\begin{equation}
\gamma \equiv \frac \omega {gt}\sqrt{\langle a^{+}a\rangle -\frac
14 (t+t^{-1})+\frac 12},  \label{ExactScaling}
\end{equation}
where we have defined $t=-\langle \left( a^{+}-a\right) ^2\rangle
$. In the bipolaron regime with strong couplings, this quantity
becomes the scaling ratio, $\gamma \rightarrow \zeta /\xi $. In
this regime, if the scaling relation \eqref{scalingEq} holds, the
value of $\gamma $ will be equal to 1. Indeed, this scaling
relation is confirmed by the exact numerics, as shown in the main text.

In the quadpolaron regime with intermediate or weak couplings, not
only the scaling relation \eqref{scalingEq} is violated but also
the relation between $\gamma $ and $\zeta /\xi $ is breaking
down, $\gamma \neq \zeta /\xi $. Nevertheless, we find that,
besides the bipolaron regime having the value $ \gamma =1$, the
quadpolaron with four strong channels of tunneling is located in
the range $\gamma <1$ and the state with decaying left-right
tunnelings ($S_{\alpha \bar \alpha }$, $S_{\beta \bar \beta }$)
falls in a range $\gamma >1$, as one can see in
Fig.\ref{fig-Sij}a-c,g-i. In this sense, according to the values
and behavior of $\gamma $, one can distinguish the quantum
states of the bipolaron, quadpolaron and their changeover,
respectively.

\subsection{Scaling relation alternatively obtained from the lowest-order expansion of the effective
potential}\label{Sect-ExpanPoten}

\begin{figure}[t]
\begin{center}
\includegraphics[width=0.9\columnwidth]{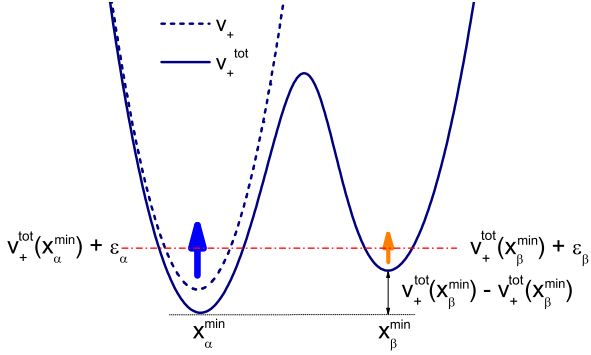}
\end{center}
\par
\vskip+0.0cm \caption{(Color online) Spin-up single-particle
effective potential, $v_+^{tot}=v_{+}+\delta  v_{+}$, in the strong
coupling regime. As in this regime within the same spin component
there is no overlap between the polaron ($\alpha$, labelled by the
blue arrow) and antipolaron ($\beta$, labelled by the orange
arrow) in the two wells, to have both finite weights for the
polaron and the antipolaron the two sub-well energies have to be
degenerate, i.e., $v_{+}^{tot}(x^{min}_\alpha)+\varepsilon_\alpha=v_{+}^{tot}(x^{min}_\beta)+\varepsilon_\beta$.
Here $x^{min}_i$ is the position of local minimum potential and
$\varepsilon_i= \xi_i  $ (scaled by $\omega/2$),
$i=\alpha,\beta$.} \label{fig-degeneracy}
\end{figure}
\begin{figure}[t]
\begin{center}
\includegraphics[width=0.8\columnwidth]{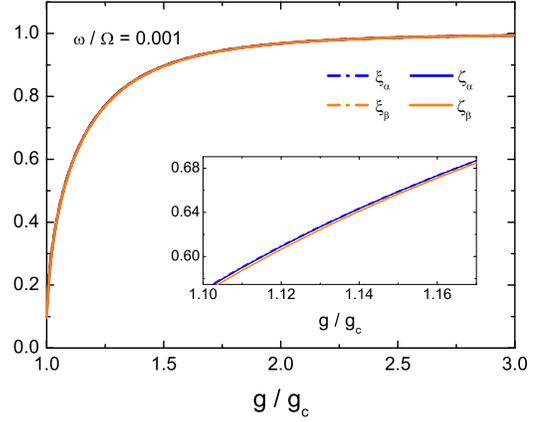}
\end{center}
\par
\caption{(Color online) Scaling relation alternatively obtained from the expansion of $v_+^{tot}$. $\omega / \Omega =0.001$ is taken. $\xi_i$ and $\zeta_i$
almost take the same values in the strong coupling regime above
the $g_c$. The inset shows their tiny differences by a zoom-in
plot. } \label{fig-scalingFromVeff}
\end{figure}

Apart from variational method on energy minimization introduced in
Appendix \ref{App-A}, an alternative way to see the
scaling relation is to investigate the effective potential. As we
discussed in the main text, the eigenequation actually can be written in
a single particle form
\begin{equation}
\frac 12\omega\left( \hat p^2+v_{\pm }^{tot}\right) \psi _{\pm}=E\psi _{\pm },
\end{equation}
where we have assumed that the particle mass $m=1$ and the total
potential is composed of the bare potential $v_{\pm }$ and an
additional effective potential $\delta v_{\pm }$ induced by the
tunneling,
\begin{equation}
v_{\pm }^{tot}=v_{\pm }+\delta v_{\pm },
\end{equation}
with
\begin{equation}
v_{\pm }= \left( x\pm g^{\prime }\right) ^2,\quad \delta
v_{\pm }=\eta \frac \Omega {\omega}\frac{\psi _{\mp }}{\psi _{\pm
}},
\end{equation}
and we have considered the ground state with $\eta =-1$. In
the strong coupling regime, the total potential exhibits an
obvious two-well structure, with a larger barrier separating the
wells, as shown in Fig. \ref {fig-degeneracy}.

In the lowest order, the two wells can be considered as a local
harmonic potential. Actually, an expansion around the local
minimum point $x_i^{\min } $ of the potential leads to
\begin{equation}
v_{+}^{tot}\cong v_{+}^{tot}(x_i^{\min })+f_i^{(1)}\left(
x-x_i^{\min }\right) +f_i^{(2)}\left( x-x_i^{\min }\right) ^2,
\end{equation}
where $x_i^{\min }=\eta _i\zeta _ig^{\prime }$ with $i=\alpha
,\beta $, and $ \eta _\alpha =-1$, $\eta _\beta =1$ respectively
for the polaron and the antipolaron. The coefficients are defined
by
\begin{equation}
f_i^{(1)}=\frac{\partial v_{+}^{tot}}{\partial x}|_{x=x_i^{\min
}},\quad f_i^{(2)}=\frac 12\frac{\partial ^2v_{+}^{tot}}{\partial
x^2}|_{x=x_i^{\min }}.
\end{equation}
First, the approximation of local harmonic potential requires
\begin{eqnarray}
\text{condition-1:} &&\quad f_i^{(1)}=0, \\
\text{condition-2:} &&\quad f_i^{(2)}=\xi _i^2.
\end{eqnarray}
The condition-1 ensures the potential minimum point at $x_i^{min}$,
while the condition-2 indicates the same renormalized frequency
$\xi_i\omega$ of the local harmonic potential as that of the
wavefunction of the harmonic oscillator. Furthermore, since effectively there is no single-particle inter-site hopping (if regarding the wells as two sites) in the presence of the large barrier in the strong coupling regime, to have finite weights for both the polaron and the
antipolaron in the single-particle effective potential the local
energies of the two wells need to be degenerate
\begin{equation}
\text{condition-3:\quad }v_{+}^{tot}(x_\alpha ^{\min
})+\varepsilon _\alpha =v_{+}^{tot}(x_\beta ^{\min })+\varepsilon
_\beta ,
\end{equation}
where
\begin{equation}
\varepsilon _i= \xi _i
\end{equation}
is the energy of the local harmonic oscillator scaled by $\omega/2$
and
\begin{equation}
v_{+}^{tot}(x_i^{\min })=v_{+}(x_i^{\min })+\delta v_{+}(x_i^{\min
})
\end{equation}
corresponds to the reference energy. Taking the variational
wavefunction (\ref {wf-1}), in the strong coupling regime we have
\begin{eqnarray}
\delta v_{+}(x_\alpha ^{\min })\doteq -\frac \Omega {\omega}
\frac{\beta \varphi
_\beta (-x_\alpha ^{\min })}{\alpha \varphi _\alpha (x_\alpha ^{\min })}%
,\nonumber \\
\delta v_{+}(x_\beta ^{\min })\doteq -\frac \Omega
{\omega}\frac{\alpha
\varphi _\alpha (-x_\beta ^{\min })}{\beta \varphi _\beta (x_\beta ^{\min })}.
\end{eqnarray}
while $\beta =\sqrt{1-\alpha ^2}$ in the strong coupling regime.

Now one can (i) control the displacement renormalization $\zeta
_i$ to satisfy condition-1 so that the linear term $f_i^{(1)}$ is
eliminated and the minimum is located at $x_i^{\min }$, (ii) tune
the frequency renormalization $\xi _i$ to fulfill condition-2 so
that both the local potential and the wavefunction
self-consistently shares the same frequency $ \xi _i\omega $,
(iii) balance the weight ratio of $\alpha /\beta $ to meet
condition-3 so that the two wells have degenerate local energies
to self-consistently guarantee the finiteness of the weights
$\alpha $ and $ \beta $. At this point, we see that the degree of
the basic deformation factors introduced for our variational
wavefunction is consistent with the minimum requirements of
self-consistence conditions.

From Conditions-1,2,3 we also refind the scaling relation as
illustrated by Fig.\ref{fig-scalingFromVeff}, which might provide
some alternative insights for the scaling relation that we
obtained from the energy minimization in last subsection. Still,
we should mention there is a small difference between the two ways,
since the above consideration from the effective potential is
based on the lowest order expansion which guarantees only the
local potential itself to be harmonic without taking care of
the energy, while the energy minimization ensures only the most
favorable energy but the effective potential $\delta v_{\pm }=\eta
\frac \Omega {\omega}\frac{\psi _{\mp }}{ \psi _{\pm }}$ includes
higher order terms beyond the harmonic potential approximation.
Despite the small difference, both approaches lead to the scaling
relation.

\subsection{Symmetry-breaking-like aspect for the bipolaron-quadpolaron quantum state changeover}

With the scaling relation at hand, it might provide some more
insight to discuss the quantum state changeover from the symmetry
point of view. Generally, for all eigenstates, the Hamiltonian has
the parity symmetry, ${\cal P}=\sigma_x(-1)^{a^{+}a}$ which involves
simultaneous reversion of the spin and the space. Specifically for
the ground state that we are focusing on in this work, one could find
extra symmetries. In fact, in the low frequency limit the photon
number vanishes for the ground state below $g_c$, as indicated
by Fig.\ref{fig-PhyQuan}b (this is more obvious for lower frequencies), so that additionally the total parity
symmetry can be decomposed into separate spin reversal symmetry
$\sigma_x$ and oscillator spatial reversal symmetry
$(-1)^{a^{+}a}$. These additional symmetries are broken beyond
$g_c$ due to the emergence of a number of photons, so that there
is a subsymmetry transition when the system goes across $g_c$.
Still, these spin and spatial subsymmetries are considered from
the weak coupling side and become less valid at finite frequencies
due to a nonvanishing photon number. Nevertheless, our finding of the scaling relation
provides compensation but from the strong coupling side.
Actually, as shown in Fig.4, the physical quantity we
proposed, $\gamma$, demonstrates an invariant behavior beyond
$g_c$, which confirms the scaling relation and thus the symmetric
aspect between the displacement and frequency renormalizations in
this regime. Note that, as shown in last subSection, in
this bipolaron regime the remaining left-left and right-right
tunnelings (i.e. polaron-antipolaron inter-tunnelings) render both the polaron and the antipolaron to have
finite weights, while to preserve finite weights as a quantum
effect in the absence of left-right tunneling channels (i.e. intra-polaron and intra-antipolaron tunnelings) the
polaron and the antipolaron have to maintain the
displacement-frequency scaling relation. In other words, this
displacement-frequency symmetry arises only in the absence of the
left-right channels, and conversely, there will be no left-right
channels if the symmetry is preserved there. Going from the
bipolaron regime to the quadpolaron regime, this symmetry will be
broken in the presence of the left-right tunneling channels. In
this sense, besides the afore-mentioned parity subsymmetry
breaking originating from the weak coupling side in the low frequency limit,
for both low frequency limit and finite frequencies there is
another hidden symmetry-breaking-like behavior in the changeover
of the two quantum states stemming from the strong coupling side.
Thus, it is interesting to see a deeper nature of the interaction that not only induces the bipolaron/quadpolaron quantum state changeover but also brings about the symmetry breaking.

\section{Physical implications and method extension to the multiple-mode case} \label{App-D}

\begin{figure}[t]
\begin{center}
\includegraphics[width=0.8\columnwidth]{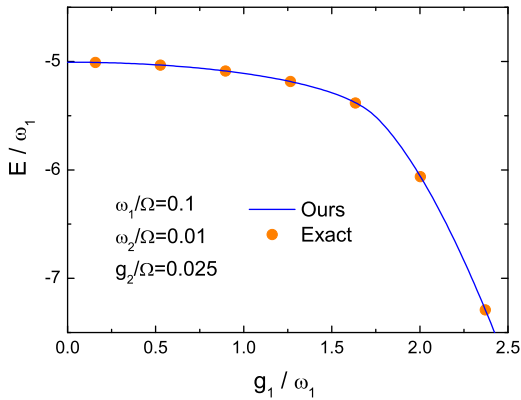}
\end{center}
\caption{(Color online) Ground-state energy as a function of $g_1$ in the two-mode case. The parameters used are $\omega_1/\Omega=0.1, \omega_2/\Omega=0.01$ and $g_2/\Omega=0.025$. Our variational method (solid line) is in good agreement with exact numerics(dots). } \label{fig-2modes}
\end{figure}

\subsection{Physical implications for the spin-boson model}

Our ground-state phase diagram obtained for the Rabi model might also provide some insights or implications for the spin-boson model \cite{Spin-Boson} which is a multiple-mode version of the Rabi model and has wide relevance to other fields, including the Kondo model \cite{RelevanceKondo} and the Ising spin chain \cite{RelevanceIsing} in condensed matter.

On the one hand, the bipolaron/quadpolaron changeover in the Rabi model can provide insights for localized/delocalized transition in the spin-boson model. In fact, the spin-boson model exhibits different behaviors in the Ohmic, super-Ohmic, and sub-Ohmic spectra, which actually have different weights of distributions for low and high frequency modes. Note that, as discussed in our work on the nature of the interaction-induced variation, the bipolaron and the quadpolaron states respectively have blockaded and enhanced left-right tunnelings, which is closely related to the situation of the localized and delocalized states involved in the spin-boson model. As indicated by our ground-state phase diagram and the obtained $g_c$ expression, the same coupling could be located in different regimes depending on whether frequency is low or high. Our ground-state phase diagram and $g_c$ expression might provide a primary reference and some insights into the different behaviours of the Ohmic, super-Ohmic, and sub-Ohmic spectra, since the distribution weights of low and high frequencies would make different contributions to blockaded and enhanced tunneling, thus affecting the competition in the quantum phase transition of the localized and delocalized states.

On the other hand, the overweighted antipolaron region might have some implication for the coherence-incoherence transition in the spin-boson model. It has been found that, within the delocalized phase of the spin-boson model, there is possibly another coherence-incoherence transition \cite{Tong2011,Spin-Boson} for which the nature is still not very clear. Interestingly, in our ground-state phase diagram of the Rabi model, within the strong-tunneling regime in the quadpolaron state, there is also an underlying particular region characterized by an unexpected overweighted antipolaron, the possible implication and relation of the overweighted antipolaron regime in the Rabi model and the coherence-incoherence transition in the spin-boson model might be worthwhile exploring.

Since in the present work our focus is on the single-mode Rabi model, we would like to leave the investigations of these possible implications for the spin-boson model to some future works. Nevertheless, in the following we provide some indication of the method and variational wavefunction.

\subsection{Method extension to the multiple-mode case}

The basic variational physical ingredients introduced in the
single-mode case also should apply for the multiple-mode case. The
treatments are readily extendable from the single-mode case. The
Hamiltonian including $M$ modes of harmonic oscillators can be
written as
\begin{equation}
H=\sum_{k=1}^M\omega _ka_k^{\dagger }a_k+\sigma
_z\sum_{k=1}^Mg_k(a_k^{\dagger }+a_k)+\frac \Omega 2\sigma _x.
\end{equation}
We propose the variational trial wavefunction as
\begin{equation}
\psi [\{x_k\}]=\alpha \prod_{k=1}^M\varphi _\alpha ^k+\beta
\prod_{k=1}^M\varphi _\beta ^k,
\end{equation}
where $\varphi _{\alpha }^k$ ($\varphi _{\beta }^k$) is the $k$'th
mode polaron (antipolaron) under the direct extension $\{\omega
,g,x,\xi _i,\zeta _i\}\rightarrow \{\omega _k,g_k,x_k,\xi
_i^k,\zeta _i^k\}$.  The energy is simply that of the single-mode
energy in \eqref{energy-1mode} with the overlap integrals replaced
by the product of all modes. We find the same accuracy as in the
single-mode case, as illustrated in Fig.\ref{fig-2modes} by an
example of the two-mode case, for which exact numerics are available
for comparison. One can also include a bias term $\epsilon \sigma
_z$ with a broken parity for $\psi _\sigma \left( x\right) $ at
different spin $\sigma$. The multiple-mode case deserves special
investigations in detail which we shall discuss in future works.



\end{document}